# Hypothesis learning in automated experiment: application to combinatorial materials libraries

Maxim Ziatdinov,[1,2,a] Yongtao Liu,[2] Anna N. Morozovska,[3] Eugene A. Eliseev,[4] Xiaohang Zhang,[5] Ichiro Takeuchi,[5] and Sergei V. Kalinin[2,b]

[1] Computational Sciences and Engineering Division, Oak Ridge National Laboratory, Oak Ridge, TN 37831

[2] Center for Nanophase Materials Sciences, Oak Ridge National Laboratory, Oak Ridge, TN 37831

[3] Institute of Physics, National Academy of Sciences of Ukraine, 46, pr. Nauky, 03028 Kyiv, Ukraine

[4] Institute for Problems of Materials Science, National Academy of Sciences of Ukraine, Krjijanovskogo 3, 03142 Kyiv, Ukraine

[5] Department of Materials Science and Engineering, University of Maryland, College Park, MD 20742

Machine learning is rapidly becoming an integral part of experimental physical discovery via automated and high-throughput synthesis, and active experiments in scattering and electron/probe microscopy. This, in turn, necessitates the development of active learning methods capable of exploring relevant parameter spaces with the smallest number of steps. Here we introduce an active learning approach based on co-navigation of the hypothesis and experimental spaces. This is realized by combining the structured Gaussian Processes containing probabilistic models of the possible system's behaviors (hypotheses) with reinforcement learning policy refinement (discovery). This approach closely resembles classical human-driven physical discovery, when several alternative hypotheses realized via models with adjustable parameters are tested during an experiment. We demonstrate this approach for exploring concentration-induced phase transitions in combinatorial libraries of Sm-doped $BiFeO_3$ using Piezoresponse Force Microscopy, but it is straightforward to extend it to higher-dimensional parameter spaces and more complex physical problems once the experimental workflow and hypothesis-generation are available.

[a] ziatdinovma@ornl.gov
[b] sergei2@ornl.gov

Recent advances in accelerated and combinatorial synthesis have allowed rapid fabrication of immense number of materials' compositions in the form of combinatorial spread libraries in pulsed laser deposition,[1-7] high-throughput microfluidic[8-12] and pipetting robot sets.[13-19] In parallel, conventional sample fabrication has been drastically accelerated via laboratory robotization and autonomous and combined human-automation workflows.[20-24] Complementary to the large-throughput synthesis are rapid characterization methods that allow establishing structure-property relations in the specific systems and in certain cases target functionalities of interest.

However, even with this outstanding progress, simple grid-based exploration of the immense compositional spaces is impractical. For example, for a quarternary alloy with the properties of interest peaking within 1% from the morphotropic phase boundary or quantum critical point, a standard grid search would require sampling $10^6$ compositions for discovery. Correspondingly, numerous strategies for active learning in such settings were proposed, using statistical learning methods to navigate compositional space. These were based on direct Bayesian optimization (BO) of the functionality of interest[25-27] or relied on the presence of cheap proxy signals to select candidates for expensive target measurements. In several cases, theoretical calculations were used to augment the BO-based strategies.[28-30]

However, fundamentally, this discovery process has been preponderantly based on the Gaussian process (GP) regression,[31, 32] a non-parametric method that learns the behavior of function of interest over relatively low-dimensional parameter space. The GP is an example of a purely data-driven method[33] aiming to interpolate the behavior of interest and its uncertainty over the specified domain based on the measurements in discrete locations. The characteristic aspect of classical GP is that the expectation value of the parameter of interest is assumed to be zero, and variability across the parameter space is captured via correlations only. The kernel function defining the degree of correlation within this parameter space is typically defined within a certain functional form and the parameters of this function are learned from the available data at each optimization/exploration step. However, the simple GP-based methods do not incorporate any physical model of behavior within the system. Consequently, they can be prone to trivial solutions, often require a fairly large number of steps to reconstruct the complicated data distributions, and are limited to low-dimensional parameter spaces.

An alternative approach for the active learning is based on the reinforcement learning (RL) methods.[34-37] The RL field has experienced rapid growth over last several years and comprises multiple subfields. However, common to simple RL algorithms is extreme data requirements, stemming from the necessity to build either a low-dimensional model of the process, or discover low-dimensional representations of the state and policies of the system. Parenthetically, it is important to note that many classical RL problems become trivial if the generative physical model of the system is known explicitly.

Recently, we have introduced the structured GP (*s*GP) approach combining the flexibility of GP models with the expressive power of the physical priors, and demonstrated significant acceleration of the active learning/optimization based on the *s*GP.[38] Here, we demonstrate that this approach can be further extended towards physics discovery via active learning of competing hypotheses, effectively combining the *s*GP and reinforcement learning. We refer to this approach as hypothesis learning. This approach is illustrated for exploring concentration-induced phase transition in Sm-doped $BiFeO_3$ using a Piezoresponse force microscopy (PFM) but can be expected to be applicable universally.

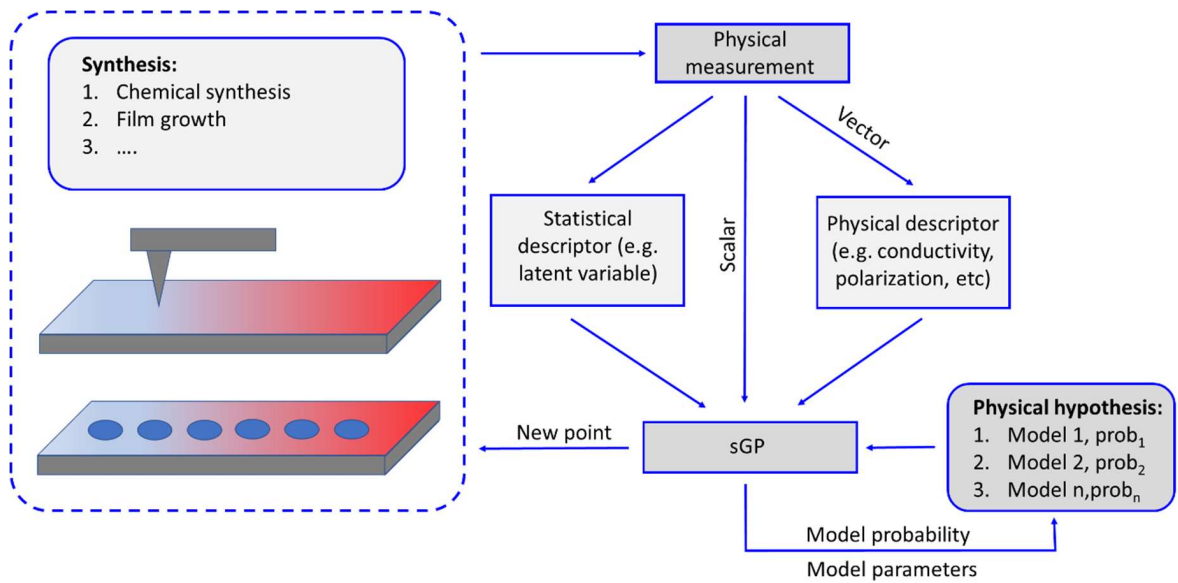

**Figure 1.** Hypothesis learning. The workflow of the scientific discovery process based on co-navigation of the experimental and hypothesis space.

The general concept of the proposed hypothesis learning approach is illustrated in Figure 1. Here, we assume a classical experimental set-up in the context of sequential materials synthesis or sequential characterization of compositional- or parameter spread systems. The examples of the former are classical wet chemical solid-state synthesis/ceramics processing, or film growth via techniques such as pulsed laser deposition, atomic layer deposition, or chemical deposition. Common to all these techniques is that the material is fabricated for a fixed set of parameters such as component ratios, stoichiometries, and deposition conditions. Jointly, these form a (relatively) low-dimensional parameter space spanning chemical and physical control variables. Once the film is grown (or material is synthesized), its properties are explored and based on the results, a new exploration point (e.g., different composition or growth condition) is chosen. Notably, a similar approach can be applied for composition spread libraries (where the composition changes across the sample plane), thickness or growth temperature gradients,[4, 39] or samples under heat, electric, or magnetic field gradients, if the local measurements are performed sequentially. These in turn can be based on the modalities of scanning probe microscopies (piezoresponse, conductive, microwave, acoustic), electron microscopy, micro-Raman imaging, or on the classical semiconductor characterization performed via microfabricated electrode arrays.

Traditionally, these synthesis-measurement cycles are performed using human intuition and past experience for selecting initial parameter settings, and often proceed via dense grid sampling of the compositional or parameter spaces. This can include characterization of the 1D compositional library, or synthesis of the epitaxial solid solution films. However, such a strategy is not efficient, since functionalities of interest are often concentrated in specific parts of the parameter space, necessitating the development of active learning methods for experimental planning, namely selection of the next experimental point based on the results of previous experiments.

In the last several years, the emergence of the automated synthesis and control systems have stimulated the implementation of the Gaussian Process-based active learning (GP-AL) and Bayesian Optimization (GP-BO) methods. By definition, GP seeks to reconstruct a scalar function $f(x)$ over $N$-dimensional parameter space $x \in R^N$ from the sparse observations $(y_i, x_i)$, $i = 1, \ldots n$, where the observations are the noisy values of the function. From the perspective of the physical sciences, the GP seeks to reconstruct a physical property of interest over the entire parameter space based on the limited number of noisy measurements of this property. To do this, the classical GP method introduces all possible functions with the zero mean over the parameter space and assumes that their probability is updated based on measurement results. The mean and dispersion of the resulting functions at each point then provide the interpolated behavior of the physical property of interest and its uncertainty over the full parameter space.

Both the predicted target property values and the corresponding uncertainty estimates can be directly used as a basis for the AL or BO, which both represent a sequential measurement protocol where the subsequent point for measuring is selected based on the previous ones. In the GP-AL, we operate in the pure exploratory mode seeking to minimize the uncertainty by selecting at each step a point associated with the largest uncertainty. In the GP-BO, the predicted values and uncertainty are combined in the acquisition function and used to guide the optimization of the (unknown) underlying function describing the property of interest. The principles and applications of the GP-BO are described in details in Ref[40].

The fundamental limitation of the GP-based AL and BO for physical sciences is that the state of the system is learned in the form of the kernel function describing the correlations between functional behavior. The latter can be chosen from a given functional form, most typically radial basis function or Matern.[41] In special cases, more complex kernel functions such as asymmetric Gaussian or spectral kernels are chosen. However, while this approach allows building a non-parametric model of system behavior, the kernel function itself is not directly linked to the physics of the system. Curiously, in most cases the mean function of GP is taken to be zero. In this manner, variability of system behavior is effectively reduced to correlation structure.

Note that the alternative approach for the analysis of physical behaviors can be found in the field of classical Bayesian inference,[42, 43] where a probabilistic model is constructed using the known functional form(s) reflecting a physical model, and some prior knowledge on model parameters.[44-46] Given the experimental observations, the posteriors of the parameter values are inferred, representing an increase of knowledge based on observations.

Here, we introduce a combined approach, referred to as hypothesis learning, integrating active discovery and Bayesian Inference, and further integrate this approach with a structured GP. Our method is summarized in Algorithm 1 and is based on the idea that during active learning a correct model of the system's behavior decreases the overall Bayesian uncertainty about the system under study. The method consists of the (very) short warm-up phase and exploration phase and assumes the existence of several measurements at randomly (or uniformly) initialized coordinates of the parameter space (we call them 'seed' measurements). The role of the warm-up phase is to produce a 'momentum' for the exploration phase. This is necessary since in the active learning setup we aim at discovering the overall data distribution with a minimal number of steps which is usually much smaller than the number of exploration steps in standard RL problems. We also must supply to the algorithm a list of probabilistic models that, based on our understanding of the system, can describe the system's behavior. These models can come as standalone parametric models with appropriate priors over their parameters or they can be wrapped into *s*GPs. The practical differences between the two approaches will be explained further in the text. It is assumed that there is one 'correct' model in the list, but we do not know which one. We also note that if the 'correct' model is absent, the flexibility of the *s*GP approach allows for convergence comparable to vanilla GP.

## Algorithm 1 (Hypothesis Learning)

**Inputs:** Initial measurements $D$, List of unmeasured points $X_*$, List of models $M$ (standalone or wrapped into $sGP$) and corresponding sample-averaged rewards $R_a$ (initialized at zeros), Exploration steps $N_{steps}$, Warm-up steps $N_{warmup}$, Reward function $R$, $\varepsilon$ in the $\varepsilon$-greedy policy.

**for** $i=1, \ldots, N_{warmup}$ **do**
   **for** each model in $M$ **do**
      Run BI to obtain posterior samples for model parameters given $D$
      Compute posterior predictive uncertainty, $\mathbb{V}[f_*]$, over $X_*$
      Store the total uncertainty, $\mathbb{V}_{tot} = \sum_{i=1}^{N=\text{len}(X_*)} \mathbb{V}[f_{*_i}]$, of the predicted function values
   **end**
   Reward the model that produced the lowest uncertainty in prediction
   Use the rewarded model to derive the next measurement point, $x_{next} = argmax(\mathbb{V}[f_*])$
   Perform measurement in $x_{next}$; update $D$ and $X_*$
**end**
Average model rewards over the warm-up steps, update $R_a$
**for** $i=1, \ldots, N_{steps}$ **do**
   Use $\varepsilon$-greedy policy to sample model from $M$ based on $R_a[i]$ or at random
   Run BI to obtain posterior samples for parameters of selected model given $D$
   Compute posterior predictive uncertainty, $\mathbb{V}[f_*]$, over $X_*$
   Reward/penalize according to $R$ ($\mathbb{V}_{tot}^i$, $\mathbb{V}_{tot}^{i-1}$), update $R_a$
   Derive the next measurement point, $x_{next} = argmax(\mathbb{V}[f_*])$
   Perform measurement in $x_{next}$; update $D$ and $X_*$
**end**

---

Once we have our 'seed' measurements and a list of models, we proceed to the warm-up phase where we conduct Bayesian inference (BI) for *each* model. In the BI, the prior, $P(\theta)$, represents our knowledge about the system before the measurement. The measurement produces the data, $D$, based on which the posterior distribution, $P(\theta|D)$, is computed via Bayes formula as

$$P(\theta|D) = \frac{P(D|\theta)P(\theta)}{P(D)} \qquad (1)$$

where $P(D|\theta)$ represents the likelihood that this data can be generated by the model with parameters $\theta$ and the denominator $P(D) = \int_\theta P(D|\theta)P(\theta)\,d\theta$ defines the space of possible outcomes. This step yields "degree of trust" in the model via the derived posterior distributions.

In practice, the BI is performed using sampling algorithms based on the Markov Chain Monte Carlo (MCMC) techniques.[47-49] Here, specifically, we are using the No-U-Turn Sampler (NUTS) algorithm as implemented in the NumPyro probabilistic programming library.[50] The posterior predictive mean and variance for a new point, $x_*$, given the measured data, $D$, are then computed as

$$P(x_*|D) = \int_\theta P(x_*|\theta)P(\theta|D)d\theta \approx \frac{1}{N}\sum_{n=1}^{N} P(x_*|\theta^n, D) = \hat{f}_*, \qquad (2a)$$

$$\mathbb{V}[f_*] = \frac{1}{N}\sum_{n=1}^{N}(f_*^n - \hat{f}_*)^2, \qquad (2b)$$

where $\theta^n \sim P(\theta|D)$ are samples drawn from the posterior. We then select a model with the lowest total (or median) uncertainty over all the unmeasured coordinates, $X_*$, and reward it according to a pre-defined reward function. The uncertainty map produced by the rewarded model is then used to select the coordinate(s) of the next measurement point the same way as in the standard AL setup. This "warm-up" procedure can be repeated multiple times ($N_{\text{warmup}}$ in Algorithm 1) to ensure that the model selection was not affected by bad initialization of the seed points. We note that conducting BI for each model is computationally costly and time-consuming (the cost and time quickly rise as new points are added) and in practice, depending on the number of models, we limit the number of warm-up steps to ~1-5.

After the warm-up phase is completed, we use a standard $\varepsilon$-greedy policy to sample only one model at each exploration step. The total uncertainty computed from the posterior of a sampled model is compared to the total uncertainty recorded in the previous step resulting in either positive (the uncertainty decreased) or negative (the uncertainty increased) reward. The next measurement point is then derived from the uncertainty map produced by the sampled model.

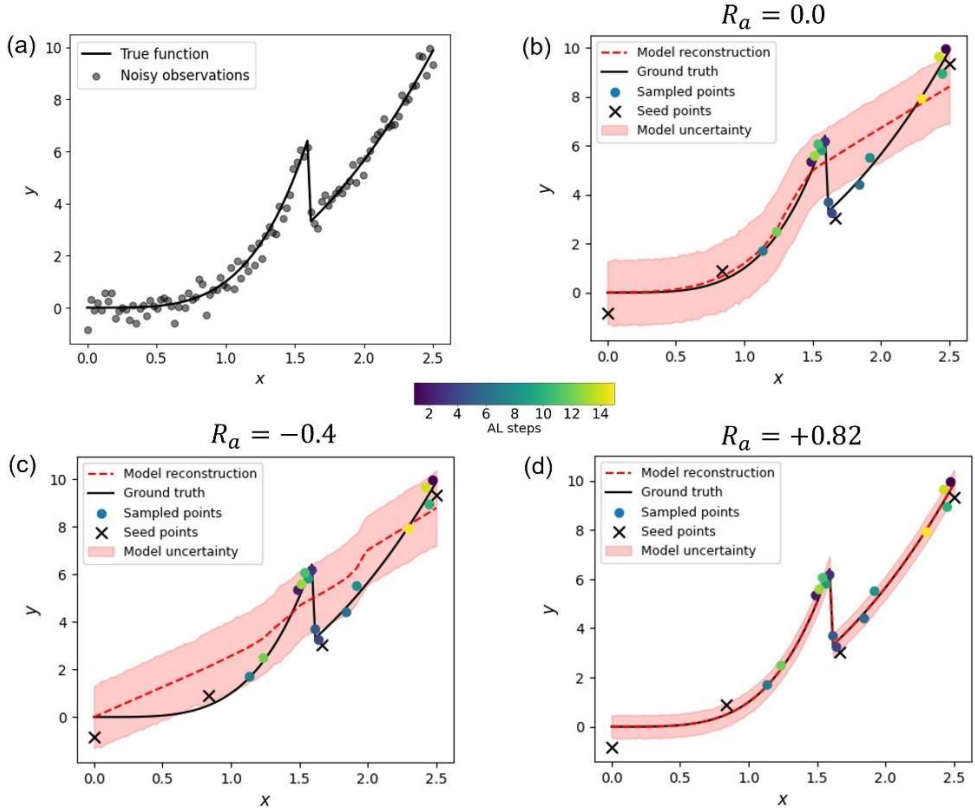

**Figure 2.** (a) Synthetic data describing a 1D system with a discontinuous "phase transition". The full data ('noisy observations') is shown but is never seen by our models. (b-d) The results of active learning after 15 steps (3 warm-up + 12 exploration) with standalone probabilistic models: model 1 (b), model 2 (e), model 3 (d), with sample-averaged rewards denoted as $R_a$.

As a practical example chosen here, we are interested in the active learning of phase diagram that has a transition between different phases. The phase transition manifests in a discontinuity of a measurable system's property, such as heat capacity. An example using synthetic data is shown in Figure 2(a) where *x*-axis is a tunable parameter (such as temperature) and *y*-axis is a system's property that is measured experimentally. However, we usually do not know where a phase transition occurs precisely, nor are we aware of the exact behavior of the property of interest in different phases. We note that using a standard GP-AL is not an optimal choice in these cases as simple GP struggles around the discontinuity point. Instead, we are going to utilize structured probabilistic models of expected system's behavior.

We hypothesize that the system of interest can be described by a piecewise function as

$$\mu(x) = \begin{cases} g_1(x), & x < x_c \\ g_2(x), & x \geq x_c \end{cases} \quad (3)$$

where $g_i(x)$ are the functions describing the system's behavior before and after the transition point $x_c$. We consider three different models based on Eq (3), namely i) a power-law behavior before transition point and linear behavior after transition point (model 1), ii) a linear behavior before and after transition point (model 2), iii) a power law behavior before and after transition point (model 3). We write them down as three standalone probabilistic models and put Bayesian priors over their parameters. Specifically, the parameter $x_c$ is modeled using a uniform prior while the rest of the parameters (slopes and power-law coefficients) are described by normal priors (for more implementation details, see the accompanying source code).

We apply the hypothesis learning (HL) algorithm to the system with a single-phase transition shown in Figure 2(a). The algorithm is initialized with four seed datapoints and our measurements at the coordinates suggested by the algorithm uncover new data points from the existing ground truth. We defined a simple reward function,

$$R\left(\mathbb{V}_{\text{tot}}^i, \mathbb{V}_{\text{tot}}^{i-1}\right) = \begin{cases} +1, & \mathbb{V}_{\text{tot}}^i < \mathbb{V}_{\text{tot}}^{i-1} \\ -1, & \mathbb{V}_{\text{tot}}^i \geq \mathbb{V}_{\text{tot}}^{i-1} \end{cases} \quad (4)$$

where $\mathbb{V}_{\text{tot}}^i$ is model's total uncertainty for the prediction over the unmeasured part of the parameter space at exploration step *i*. Here, the $\varepsilon$ parameter in the $\varepsilon$-greedy policy was set to 0.4 in all experiments, reflecting the need for exploration over small (tens of data points) number of attempts. However, similar to classical RL methods, smaller $\varepsilon$ values and annealing of $\varepsilon$ can be trivially implemented. Furthermore, the exploration policy can be based on the Thompson sampling of the posteriors or use policy embeddings. Here, given the small number of theoretical models, we did not explore these strategies.

Figure 2(b-d) shows the result of the Bayesian fit over the entire parameter space for the three models after running HL for 15 steps. The model that received the highest reward (i.e., is favored by our algorithm) clearly provided the best fit. Hence, we were able both to learn a correct data distribution with a small number of sparse measurements while also identifying a correct model that describes the system's behavior.

The evolution of the total uncertainty during the exploration for a single warm-up step and for three warm-up steps are shown in Figure 3 (a) and 3 (b), respectively, for the larger number of exploration steps. Again, in both cases, the algorithm favored the third model where the system's behavior before and after the transition point is described by the power-law functions. To confirm the reproducibility and robustness of our algorithm, we ran it for 30 different initializations. The histograms with averaged rewards received by three models are shown in Figure 3 (c) and 3 (d) for a single warm-up step and for three warm-up steps, respectively. The algorithm showed a good performance in both cases although the delineation of the correct model (model 3) was better when we ran the warm-up procedure three times.

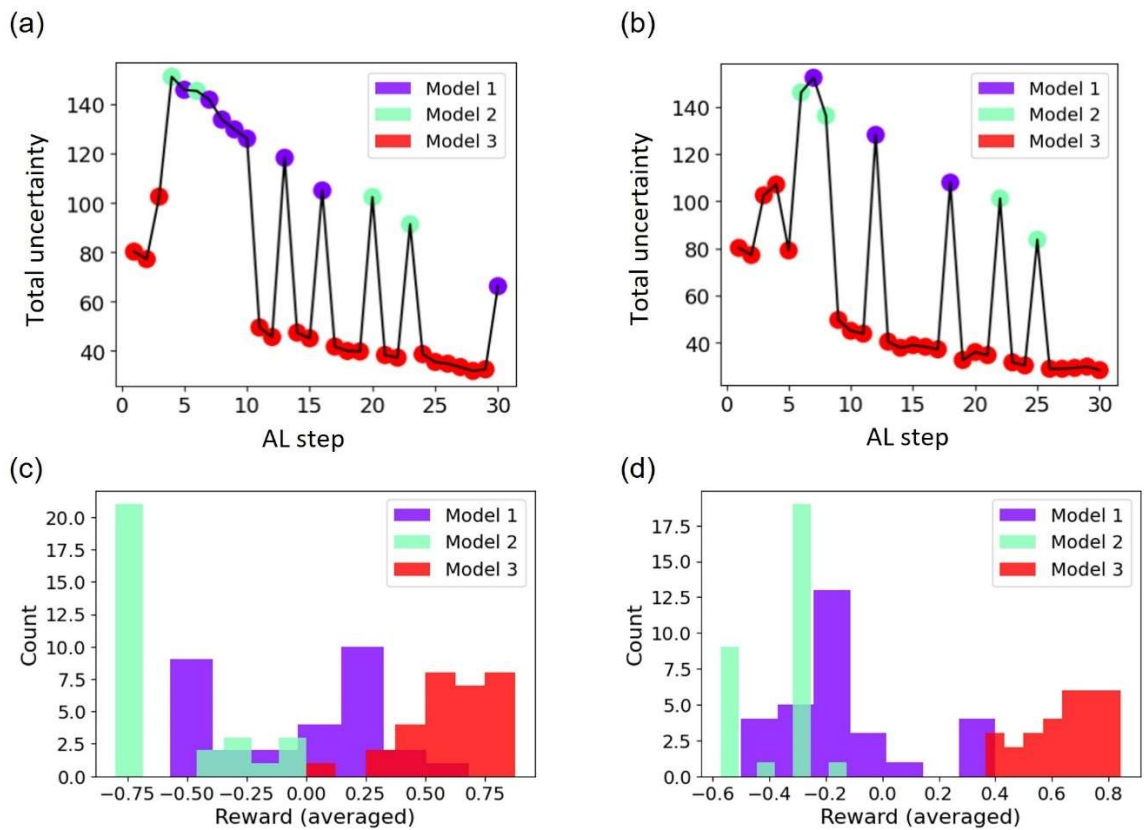

**Figure 3.** Hypothesis-driven active learning with standalone structured probabilistic models. (a, b) Evolution of the total uncertainty during the active learning with standalone probabilistic models with one 'warm-up' step (a) and with three 'warm-up' steps (b). See Algorithm 1 for the definition of the 'warm-up' steps.' At each exploration step, the model is selected according to ε-greedy policy with ε=0.4. For the warm-up steps, only the model that produced the lowest uncertainty is shown. (c, d) Statistics of the sample-averaged rewards for 30 different random initializations of the four seed measurements for one 'warm-up' step (c) and for three 'warm-up' steps (d).

One disadvantage of the current approach is that it fails when the list of probabilistic models does not contain the 'correct' model, as one can quickly conclude from Fig. 2(bc, ). At the same time, most theoretical models describe the experimental reality only to a certain approximation, and the complete/correct model may simply be not available.

To address this issue, we utilize a structured Gaussian process (sGP) method where a fully Bayesian GP is augmented by a structured probabilistic model of expected system's behavior.[38] Recall that in standard GP, we place a multivariate normal (MVN) prior over a target function $f$ such that $f \sim MVN(m(x), K(x, x'))$ and $y = f(x) + \varepsilon$, where $\varepsilon$ is a normally distributed noise. The prior mean function $m$ is usually set to 0 and the GP is completely defined by its kernel function $K$. In the sGP, we replace the GP's constant prior mean function with one of the probabilistic models defined earlier. The parameters of the GP's new probabilistic mean function $m$ and those of the GP's kernel $K$ (here chosen to be Matern[41]) are inferred simultaneously via NUTS. Hence, the posterior predictive uncertainty associated with the parameters of $s$GP's mean function is automatically taken into account in our active learning setup and provides additional control via the choice of the model's priors. The $s$GP's probabilistic mean function is expected to capture general or even partial trends in the data, but it does not have to be precise. Note that we replace standalone models $M$ in the Algorithm 1 with $s$GPs containing the same models without changing any other part of the algorithm.

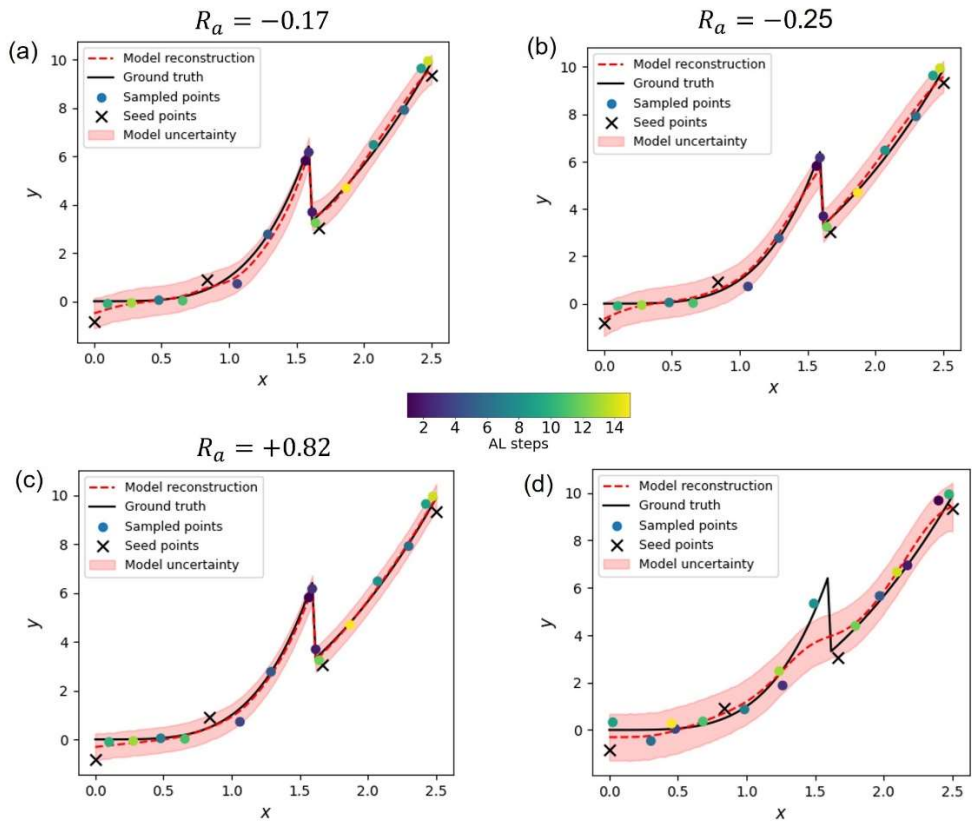

**Figure 4.** Hypothesis-driven active learning with structured Gaussian processes ($s$GPs). (a-c) The results of active learning after 15 steps (3 warm-up + 12 exploration) with $s$GPs augmented by three different structured probabilistic models (same settings as in Fig. 2b-d), with sample-averaged rewards denoted as $R_a$. Results of the active learning with vanilla GP (i.e., with prior mean function set to 0) is shown in (d) for comparison.

Figures 4 and 5 show the $s$GP results for the same set of conditions as used for the standalone probabilistic models in Fig. 2 and 3. In the $s$GP, the flexibility of the GP kernel can partially correct for the wrong model choice. This offers an advantage for the reconstruction from sparse data but also introduces a potential hurdle for learning the correct model in our setup, since even with a wrong model the total uncertainty will likely keep decreasing. Therefore, a bad initialization may result in being "locked-up" with an incorrect model (Fig. 5a).

Fortunately, this can be easily addressed by performing several warm-up steps (Fig. 5b) where we compare the absolute values of the total (or median) uncertainties between different $s$GP models. Indeed, as one can see from the histograms of sample-averaged rewards for 30 different initializations in Fig. 5c and 5d, the increased number of warmup steps allows for a more robust identification of the correct model. The Bayesian fits over the entire parameter space for the three models after running the HL for 15 steps demonstrate that even with 'wrong' models, the *s*GP can still recover the data distribution reasonably well (Fig. 4a, b). At the same time, the best reconstruction was produced by *s*GP with the third model as its prior mean function (Fig. 4c) that received the largest reward (0.82), in agreement with the results of the standalone probabilistic models and the known ground truth. For comparison, we also showed the results of the active learning with a vanilla GP (i.e., GP with a prior mean function set to 0), which completely failed around the discontinuity point (Fig. 4d). In the Supplementary Note I, we describe the application of *s*GP to several more systems with discontinuous behavior, including a scenario where none of the proposed models correctly describes a behavior of the full system.

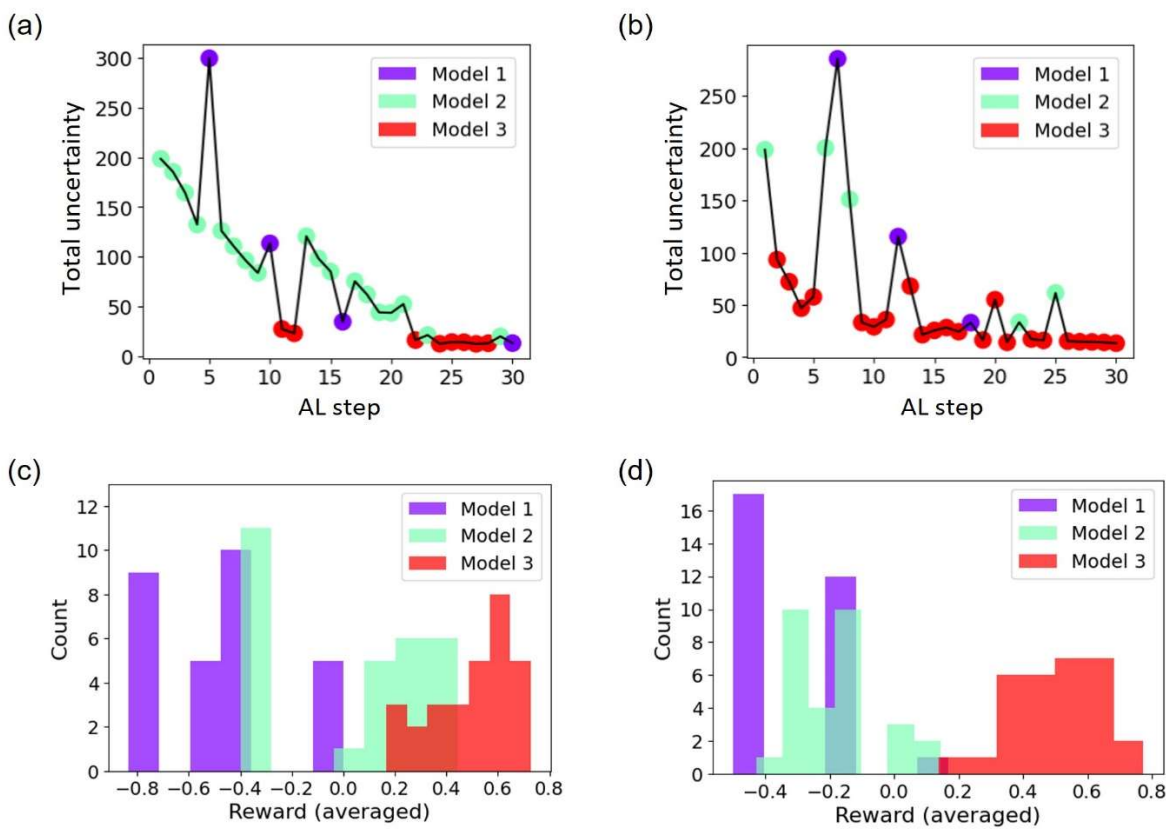

**Figure 5.** Hypothesis-driven active learning with structured Gaussian processes (*s*GPs). (a, b) Evolution of the total uncertainty during the active learning with *s*GPs augmented by three different structured probabilistic models (the models are the same as in Fig. 3) with one 'warm-up' step (a) and with three 'warm-up' steps (b). See Algorithm 1 for the definition of the 'warm-up steps.' At each exploration step, the model is selected according to $\varepsilon$-greedy policy with $\varepsilon=0.4$. For the warm-up steps, only the model that produced the lowest uncertainty is shown. (c, d) Statistics of the sample-averaged rewards for 30 different random initializations of the four seed measurements for one 'warm-up' step (c) and for three 'warm-up' steps (d)

Having confirmed the performance of our approach on synthetic data, we move to the real-time PFM experiments on the combinatorial library of Sm-doped BiFeO$_3$, as shown in Figure 6.

This material system has been extensively studied in the context of the evolution of ferroelectric properties, with the pure BiFeO$_3$ being rhombohedral ferroelectric materials and 20% Sm-doped BiFeO$_3$ being the orthorhombic non-ferroelectric.[51] The intermediate phases at the boundary between crystallographically-incompatible phases are known to exhibit complex nanoscale ordering patterns and can be associated with enhanced electromechanical responses.[52] However, these behaviors are strongly dependent on the mechanical and electrical boundary conditions,[53-55] resulting in strong variability of behaviors between bulk ceramics and thin films. Generally, the mechanisms describing the emergence of these properties are poorly understood and are of continuous interest for physical community.[56-62] The compositional spread library hence encodes the full constant-temperature cross-section of the phase diagram in a given concentration range. Previously, we have extensively explored the property of this combinatorial library using high-resolution electron microscopy.[46, 63]

Here, we have explored the mesoscale electromechanical responses in this material system using PFM experiments.[64-67] This technique is based on the detection of the local electromechanical response induced by application of small amplitude (~1 Vac) periodic bias (100-300 kHz) to the scanning probe. The detection of the response as a function of slowly (~Hz) varying large amplitude (3-10 V) waveforms allows measuring the local hysteresis loops.[68-72] The shape of the hysteresis loop is closely linked to tip-induced domain dynamics,[73, 74] and thus represents local polarization switching mechanisms.[75, 76] Here, we use the switching spectroscopy PFM approach, where hysteresis loops are measured over the rectangular grid of points, offering better statistics and information on the spatial variability of switching behavior.[77, 78] Important for subsequent discussion is that in linear approximation, the PFM signal is independent on tip-surface contact radius and hence surface topography, making PFM signal quantitative.[79-82] Shown in Figure 6a are band excitation PFM results showing the sample morphology and domain structures from three representative locations with varying Sm concentration, along with the picture of the Sm concentration gradient sample in Figure 6b. Figure 6c shows hysteresis loops from six locations corresponding to various Sm concentration, indicating the dependence of switching behavior on Sm concentration.

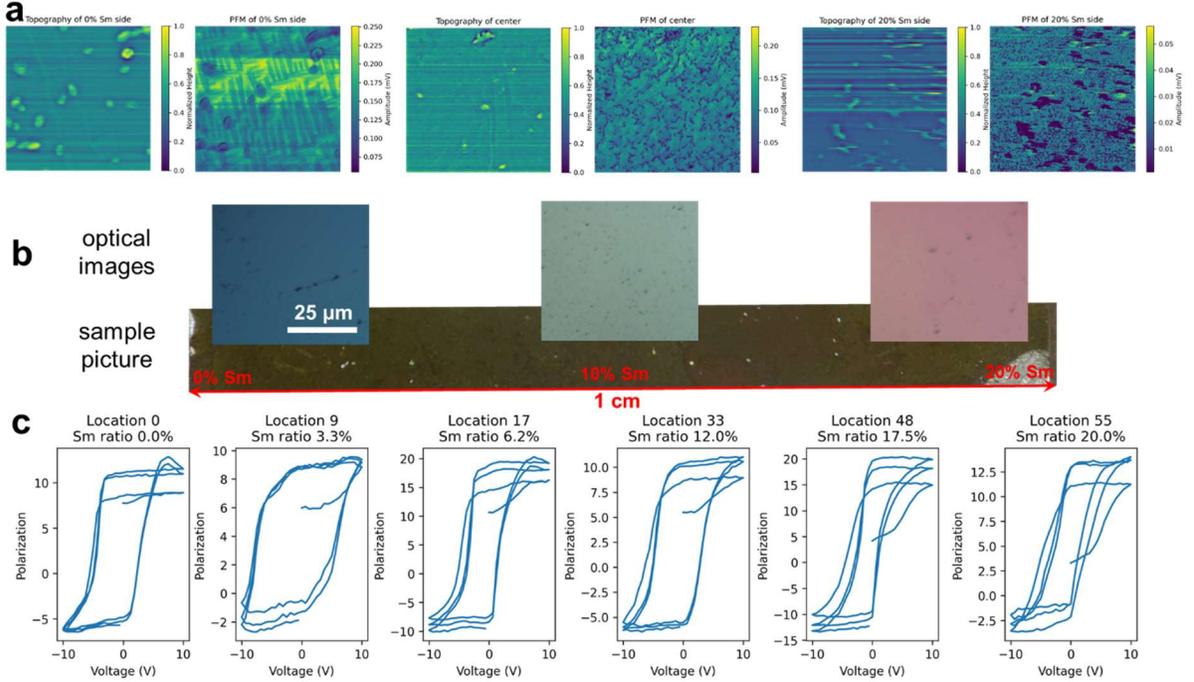

**Figure 6.** Piezoresponse force microscopy (PFM) results of Sm doped BFO sample. (a) Topography and corresponding band excitation PFM amplitude images from three representative locations: 0% Sm side, center, and 20% Sm side, which correspond to the three optical images in (b), respectively. (b) Picture of the Sm-BFO sample, which has a gradually increased Sm ratio from the left side to the right side. The optical images under a Cypher microscopy optical camera show different color due to Sm ratio variation, this color gradient can also be observed by naked eyes; however, note that the color shown by the microscope camera deviates due to optical path effects and illumination conditions. (c). Hysteresis loops from six selected locations obtained in Band excitation piezoresponse spectroscopy measurements; note that 56 locations representing gradual Sm concentration change were uniformly marked along the sample.

To determine the possible models of the hysteresis loop behavior across the compositional phase transition, we develop a simple Ginburg-Landau based model. Here, we assume that piezoresponse (PR) loops are determined by the ferroelectric polarization reversal via piezoelectric coupling as:

$$PR_3(V) \cong PR_0 + d_{33}^{eff} P_3(V), \quad (5a)$$

$$\Gamma \frac{\partial P_3}{\partial t} + \alpha(T,x)P_3 + \beta P_3^3 + \gamma P_3^5 + \delta \ln\left(\frac{P_0+P_3}{P_0-P_3}\right) = E_3(t), \quad (5b)$$

where $d_{33}^{eff}$ is an effective piezocoefficient related to a pure BiFeO$_3$. Applied voltage is $V = V_{DC}(t) + V_{AC} \sin(\omega t)$, where $V_{DC}(t)$ is a slow-varying amplitude of the voltage applied to the tip. The acting field $E_3(t) = -\frac{\varepsilon_d V_{DC}(t)}{\varepsilon_d h + \lambda \varepsilon_{33}^b}$, where $h$ is the BiFeO$_3$ thickness, and we assume that an ultra-thin gap of width $\lambda$ separates the BiFeO$_3$ surface from the tip. Constant $\varepsilon_{33}^b$ is a relative background permittivity of BFO, $\varepsilon_{ij}^b \leq 10$, and $\varepsilon_d \sim (1-10)$ is a relative permittivity in the gap.[83]

The coefficient $\alpha(T, x)$ linearly depends on the Sm concentration:[84, 85]

$$\alpha(T, x) = \alpha_T(T - T_c(x)), \qquad T_c(x) = T_C\left(1 - \frac{x}{x_{cr}}\right), \tag{6}$$

where $T$ is the temperature, $T_C$ is a Curie temperature of a pure BiFeO$_3$, and $x$ is the relative content of rare-earth Sm atoms. The critical concentration $x_{cr}$ is a fitting parameter. Note that Sm doping also affects the parameters $PR_0, \beta, \gamma$ and $\delta$, and the effect can be nonlinear and complex. Only the shift $T_c(x)$ contains a linear component.

Several cases of possible materials behaviors are analyzed in the Supplementary Note II. Here, we considered the two possible models that relate the measured loop area $S$ to the dopant concentration $x$. The first model is defined as

$$S = \begin{cases} S_0\left(1 - \frac{x}{x_0}\right)^2 + C, & x \leq x_c, \\ C, & x > x_c \end{cases} \tag{7}$$

where $x_c$ is the transition point. This model does not have a discontinuous jump at the transition point. The second model is defined as

$$S = \begin{cases} S_0\left(1 - \frac{x}{x_0}\right)^{\frac{5}{4}} + C_0, & x \leq x_c, \\ C_1, & x > x_c \end{cases} \tag{8}$$

and is characterized by a discontinuity at $x_c$. Note that this analysis is thermodynamic in nature and, in particular assumes structurally homogeneous phases. As such, it does not account for potential formation of morphotropic phases with enhanced electromechanical responses.

We note that, unlike in the examples on synthetic data, we do not know the true function in the actual experiment. We, therefore, cannot exclude the possibility that none of these models is correct. We place a uniform prior over the transition point location and log-normal priors on all the remaining parameters of the models as well as GP kernel hyperparameters and do not change them during the experiment.

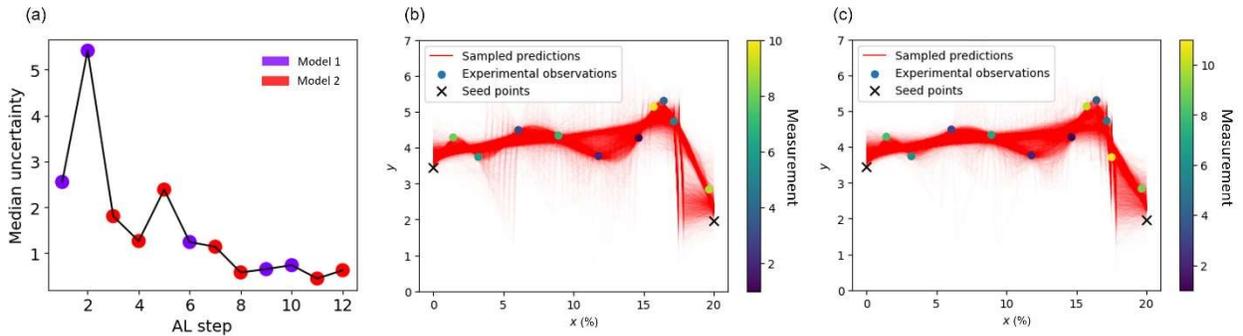

**Figure 7.** Application of Algorithm 1 with two different $s$GP models to real experiment. (a) Evolution of the uncertainty during the active learning with $s$GPs. (b, c) $s$GP predictions sampled from the posterior distribution at step 11 (b) and 12 (c), before performing a new measurement.

The experiment starts by measuring two extreme points on the concentration axis. These seed points are used to initialize Algorithm 1 with two $s$GP models. We used five warmup states and seven exploration steps. The median uncertainty as a function of the active learning steps is shown in Fig. 7a. At the end of the exploration, the first model received a cumulative sample-averaged reward of 0.13. The second model received a reward of 0.56. This suggests that the second model is better at describing the current system. However, the second model experienced significant difficulties in convergence at each exploration step, as judged by the effective number of MCMC samples and the Gelman-Rubin convergence statistics (see Supplementary Note III). The convergence problems indicate that, even though the second model received a larger reward, it may not be the correct model for describing the system's behavior (see also a discussion in the Supplementary Note I for synthetic data). To get a better understanding, we examined the *s*GP predictions sampled from the posterior distribution at the two last steps of the exploration. Shown in Fig. 7b and 7c are the sampled *s*GP predictions together with the experimental values measured in the points suggested by the algorithm. We can see there's a discontinuity-like behavior at $x \sim 17\%$, which explains why the algorithm favored the second model, preceded by a prominent peak-like feature. The latter was not a part of any prior model, which explains the problems with convergence. The inconsistency between the prior model and observations resulted in larger uncertainty around the transition region, allowing us to quickly discover a potentially new behavior. This discovery, in turn, will require the development of an improved model of the system's behavior to account for the peak before the transition, i.e., the behavior associated with the morphotropic behavior.

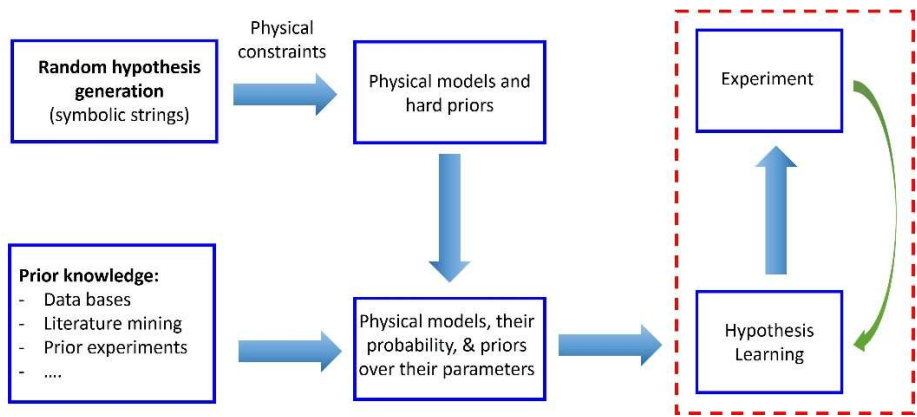

**Figure 8.** The "automated scientist" concept. Here, the initial hypothesis space is formed as a random symbolic sequence up to a given length. The hard physical constraints (dimensionality, conservation laws, symmetry) are used to select the physical hypothesis and provide hard priors (e.g., non-negativity). The prior "soft" knowledge is used to ascribe relevant probability and form parameter priors. These models form inputs to the (active) hypothesis learning algorithm. Here, the part in red dashed rectangular corresponds to Figure 1.

We further consider the opportunity to combine the hypothesis learning algorithm with the automated hypothesis generation, as illustrated in Figure 8. Over the last decade, several notable frameworks have been proposed to discover of the laws and functional expressions from the experimental data. They include Eureka originally developed for the dynamic data[86] and more recent frameworks such as PySindy,[87] PySR,[88] SISSO[89], and many others. The overview of symbolic regression methods in materials science is given in Ref [90]. These methods operate with the static data, but generally allow the generation and simplification of the symbolic functional form consistent with the observations. Recent work illustrates how this approach can be generalized to incorporate physical laws.[91] Other developments in this area includes frameworks such as Feynman AI.[92]

We argue that the same approach can be extended for experimental hypothesis learning, interfacing hypothesis generation with experiment. Clearly, the hypothesis spaces formed by all possible symbolic expressions are intractable for experimental verification, as are all physically possible models. Hence, an additional stage will be down the selection of the hypotheses using the prior knowledge derived from domain expertise or literature mining.[93-96] In the context of HL, this includes assigning probabilities to the possible theories and identifying the priors on the relevant parameters. The formed list of the hypotheses and corresponding priors can be used as an input for the HL, substituting the model list in Fig. 1. It is important to note that this approach dynamically allows further matching the degree of compression of the hypothesis space and expected experimental budget. For example, when the experiments are high throughput, and fast, hypothesis space can be expanded, whereas for expensive experiments, the effort in down selecting hypotheses is justified. Similarly, hypothesis selection can be based on different selection strategies, varying degrees of risk, level of assumption, i.e., within its own optimization strategies. Overall, this approach combines hypotheses generation and testing and allows constraining the experimental budget.

Finally, it is important to note that the hypotheses generation is not necessarily limited to symbolic models and can also include, for example, the lattice Hamiltonians and force field models. However, in this case, the inverse problems are considerably more complex and are addressed only for special cases.[97, 98] Similarly, this will necessitate transition from discrete models with Bernoulli distribution to more complex representations such as continuous embeddings, etc.

To summarize, we introduced an active learning approach based on co-navigation of hypothesis and experimental spaces via a combination of fully Bayesian structured Gaussian Processes with reinforcement learning policy refinement. This approach closely resembles classical human-driven physical discovery, when multiple alternative hypotheses realized via models with adjustable parameters are tested during the experiment. We demonstrated this method for exploring concentration-induced phase transitions in the combinatorial libraries of Sm-doped $BiFeO_3$ using PFM measurements.

Looking forward, we believe that the proposed approach makes a strong case for the synergistic combination between scanning probe and other imaging methods with combinatorial material science. While traditionally combinatorial studies have been limited by the need to quantify materials structure and functionality across the libraries, t recent advances in the focused X-Ray methods and SPM offer a solution. At the same time, the proposed hypothesis-driven Gaussian Processes framework developed here further allows incorporation and selection between physical models. Note that while the approach is implemented here for the 1D case, it is straightforward to extend it to higher-dimensional parameter spaces and more complex physical problems.

**Acknowledgements:** This effort (M.Z.) was performed and supported at Oak Ridge National Laboratory's Center for Nanophase Materials Sciences (CNMS), a U.S. Department of Energy, Office of Science User Facility. Y.L. and S.V.K. was supported as part of the center for 3D Ferroelectric Microelectronics (3DFeM), an Energy Frontier Research Center funded by the U.S. Department of Energy (DOE), Office of Science, Basic Energy Sciences under Award Number DE-SC0021118. A.N.M. was supported by the National Academy of Sciences of Ukraine (the Target Program of Basic Research of the National Academy of Sciences of Ukraine "Prospective basic research and innovative development of nanomaterials and nanotechnologies for 2020 - 2024", Project № 1/20-H, state registration number: 0120U102306) and received funding from the European Union's Horizon 2020 research and innovation programme under the Marie Skłodowska-Curie grant agreement No 778070. The work at the University of Maryland was supported in part by the National Institute of Standards and Technology Cooperative Agreement 70NANB17H301. The authors gratefully acknowledge Dr. Bobby Sumpter for careful reading and editing of the manuscript.

**Experimental details**

A Sm-doped BiFeO$_3$ (Sm-BFO) thin film, grown on SrTiO$_3$ substrate, with Sm concentration gradually increases from 0% to 20% along the sample was used as a model system. The detailed growth conditions and characterization are described in previous publications.[46, 63]

Band excitation piezoresponse spectroscopy (BEPS) measurements were performed on an Oxford Instrument Asylum Research MFP3D atomic force microscopy system. A National Instruments DAQ card and a LabView framework are equipped for band excitation measurement. All measurements were carried out using Budget Sensor Multi75E-G Cr/Pt coated AFM probes (~3 N/m).

56 locations were marked on the side of sample, which correspond to the Sm concentration spanning from 0% to 20%. Band excitation piezoresponse spectroscopy measurements were performed on the locations determined by the hypothesis learning algorithm.

**Code availability**
Code is available without restrictions at https://github.com/ziatdinovmax/hypoAL

# Supplementary Materials

**I. Application of hypoAL for active learning of different discontinuous functions.**

Here we illustrate application of hypoAL (Algorithm 1) to two different discontinuous functions. In the first case, there is one correct model in the input list $M$. In the second case, none of the models in the input list is correct.

**I.1.** In the Supplementary Figure 1, we show data generated by the discontinuous function of the form:

$$f(x) = \begin{cases} A + B/(x - x_c), & x < x_c \\ C/(x - x_c), & x \geq x_c \end{cases} \quad (S1)$$

The noisy observations generated by this function together with the 'true function' are shown in Supplementary Figure 1.

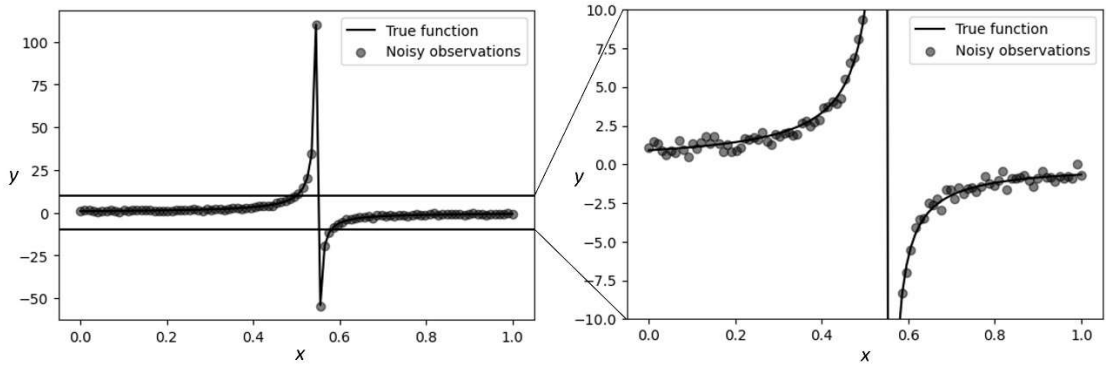

**Supplementary Figure 1.** The noisy observations of discontinuous function in Eq. S1

Same as in the main text example, the full data ("noisy observations") in Supplementary Figure 1 is never seen by the algorithm. Instead, we start with four 'seed' measurement to initialize the algorithm. The algorithm is provided with the two possible models of system's behavior, wrapped into structured Gaussian processes. The first model is a Lorentzian peak function of the form $A + B/\sqrt{(x - x_c)^2 + C^2}$, with a *Uniform(0, 1)* prior on $x_c$ and *Normal (0, 1)* priors on $A$, $B$, and $C$ parameters. The second model is the one from Eq. S1, with the same priors on its parameters.

The hypoAL results after 20 steps (3 warm-up + 17 exploration) with the same settings as used in the main text are shown in Supplementary figure 1, together with the reward values for each model.

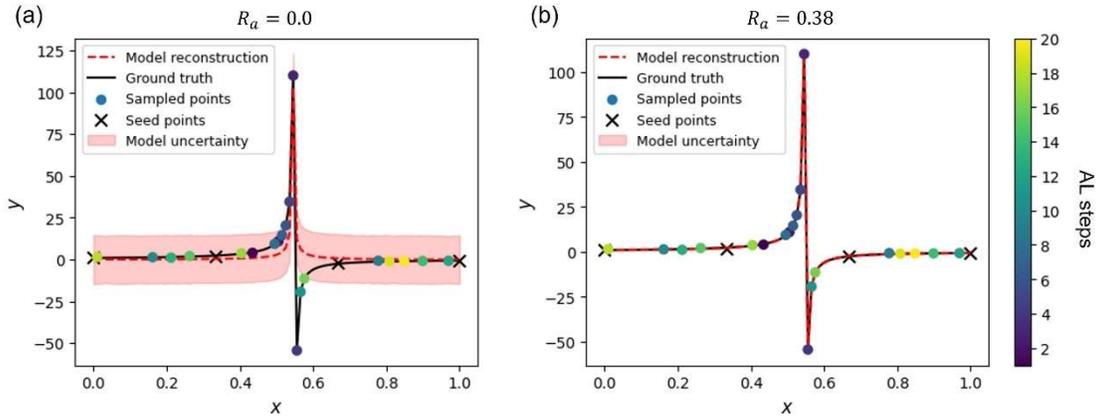

**Supplementary Figure 2.** Results of the hypotheses-driven active learning after 20 steps (3 warm-up + 17 exploration) with $s$GPs augmented by two different probabilistic models, with sample-averaged rewards denoted as $R_a$.

**I.2.** In Supplementary Figure 3, we show noisy observations of a discontinuous function which, unlike the one in the main text, has two transition points. We are going to use the same HypoAL settings with the same models, wrapped into structured Gaussian process, as in the main text. This means that none of the models is correct (as they all assume only a single transition).

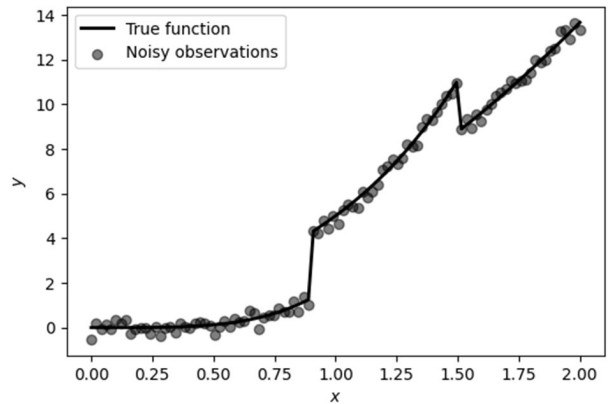

**Supplementary Figure 3.** The noisy observations of discontinuous function with two "phase transitions".

The hypoAL results are shown in Supplementary Figure 4. The first two models that assume a linear behavior after the transition point received high positive rewards whereas the third model that assumes a power law behavior before and after the transition point received a negative reward. Although none of the models is correct, the preference for the first two models can be easily understood since it is easier to approximate the behavior after the first transition point with a linear behavior than a power law one. In both cases, the flexibility of GP kernel partially compensates for the incorrect model choice. We note that the models with high rewards described correctly one of the transitions and provided a hint for the existence of the second transition.

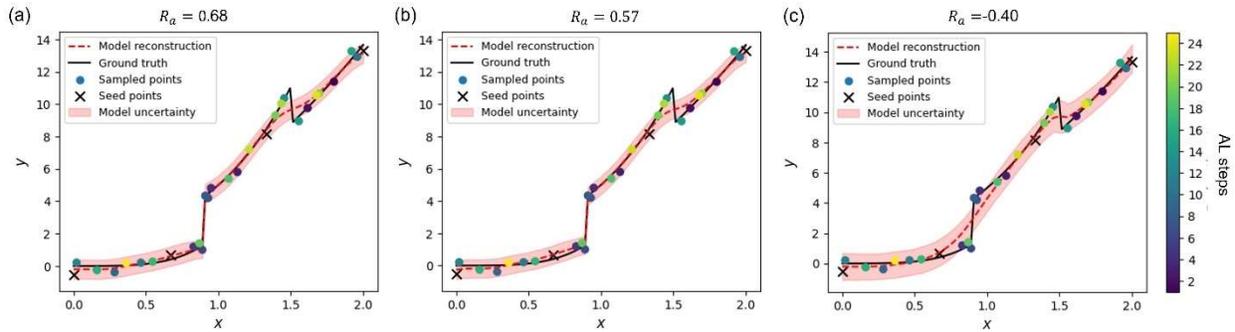

**Supplementary Figure 4.** Results of the hypotheses-driven active learning after 25 steps (3 warm-up + 12 exploration) with *s*GPs augmented by the same probabilistic models as in the main text, with sample-averaged rewards denoted as $R_a$.

In the example above, the correct piece of information in all the models was the presence of a transition point. The obvious question is: how the hypoAL algorithm distinguishes between a model that is only partially correct and a model that is completely wrong? To test this, we run hypoAL using two models: the third model in the example above and a wrong model of the form $ax^2 + b$ with *Normal(0, 1)* priors over its parameters.

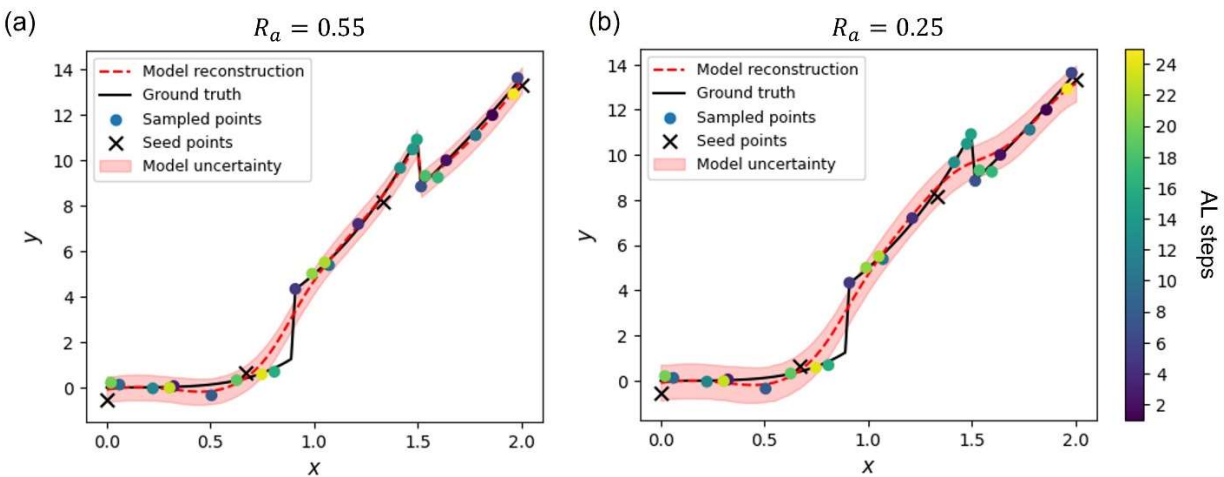

**Supplementary Figure 5.** Results of the hypotheses-driven active learning after 25 steps (3 warm-up + 12 exploration) with *s*GPs augmented by a partially correct model (a) and by a wrong model (b), with sample-averaged rewards denoted as $R_a$.

The results shown in Supplementary Figure 5 clearly demonstrate that when presented with a wrong model (Fig. S5b) and a partially correct model (Fig. S5a), the algorithm chooses the partially correct model. At the same time, a model that is only partially correct typically results in a slow MCMC convergence at each step. In this case, the MCMC diagnostics may show a low effective number of samples (n_eff) and high values (>1.05) of Gelman-Rubin criteria (r_hat), as demonstrated below for the final AL step that performed Bayesian inference on the first model parameters using NUTS with 5000 samples:

|         | n_eff  | r_hat |
|---------|--------|-------|
| beta1   | 48.07  | 1.03  |
| beta2   | 35.36  | 1.03  |
| k_length| 31.13  | 1.09  |
| k_scale | 70.95  | 1.05  |
| noise   | 58.99  | 1.00  |
| t       | 293.52 | 1.00  |

Such diagnostics can serve as an indicator that the model needs to be updated.

**II. Phenomenological models of the local piezoresponse formation in $Bi_{1-x}Sm_xFeO_3$**

One of the most widely used model for the formation of vertical PR loops is the semi-empirical equation of Landau type, which reflects the fact that PR loops are determined by the ferroelectric polarization reversal via piezoelectric coupling, rather than by electrostrictive coupling as it follows from the classical Landau theory. The mathematical formulation of the above statement is:

$$PR_3(V) \cong PR_0 + d_{33}^{eff} P_3(V), \qquad (1a)$$

$$\Gamma \frac{\partial P_3}{\partial t} + \alpha(T,x) P_3 + \beta P_3^3 + \gamma P_3^5 + \delta \ln\left(\frac{P_0 + P_3}{P_0 - P_3}\right) = E_3(t), \qquad (1b)$$

where $d_{33}^{eff}$ is an effective piezocoefficient related to a pure $BiFeO_3$. Applied voltage is $V = V_{DC}(t) + V_{AC}\sin(\omega t)$, where $V_{DC}(t)$ is a slow-varying amplitude of the voltage applied to the tip. The acting field $E_3(t) = -\frac{\varepsilon_d V_{DC}(t)}{\varepsilon_d h + \lambda \varepsilon_{33}^b}$, where $h$ is the $BiFeO_3$ thickness, and we assume that an ultra-thin gap of width $\lambda$ separates the $BiFeO_3$ surface from the tip. Constant $\varepsilon_{33}^b$ is a relative background permittivity of BFO, $\varepsilon_{ij}^b \leq 10$, and $\varepsilon_d \sim (1-10)$ is a relative permittivity in the gap.[82]

The coefficient $\alpha(T,x)$ linearly depends on the Sm concentration:[83, 84]

$$\alpha(T,x) = \alpha_T (T - T_c(x)), \qquad T_c(x) = T_C\left(1 - \frac{x}{x_{cr}}\right), \qquad (2)$$

where $T$ is the temperature, $T_C$ is a Curie temperature of a pure $BiFeO_3$, and $x$ is the relative content of rare-earth Sm atoms. The critical concentration $x_{cr}$ is a fitting parameter. Note that Sm doping affect also on the parameters $PR_0$, $\beta$, $\gamma$ and $\delta$, but the effect can be nonlinear and complex. Only the shift $T_c(x)$ contains a linear component.

**Case I.** The coefficients $\beta > 0$ and $\gamma = \delta = 0$ for the second order ferroelectrics of the displacement type For the case the width of the PR loop is proportional to $E_c = \frac{2}{3\sqrt{3}}\sqrt{-\frac{\alpha^3}{\beta}}$, and the loop height is proportional to the spontaneous polarization $P_S = \sqrt{-\frac{\alpha}{\beta}}$. The PR loop area $S$, which has a quasi-rectangular shape in a quasi-static case, is roughly proportional to their product, but has sense only when $\alpha < 0$. So:

$$S(T,x) = \begin{cases} S_0 \left(\frac{T}{T_C} - \left(1 - \frac{x}{x_{cr}}\right)\right)^2, & T \leq T_C\left(1 - \frac{x}{x_{cr}}\right), \\ 0, & T \geq T_C\left(1 - \frac{x}{x_{cr}}\right). \end{cases} \qquad (3a)$$

Here $S_0$ and $x_{cr}$ are fitting parameters, and $T_C$ is tabulated.

For the purposes of the implementation in the code, we can introduce $x_0$ as the concentration for which phase transition happens at room temperature, $T_R = T_C\left(1 - \frac{x}{x_0}\right)$. With this re-designation Eq.(3a) can be rewritten as:

$$S(T_R, x) = \begin{cases} S_R\left(1 - \frac{x}{x_0}\right)^2, & x \leq x_0, \\ 0, & x > x_0. \end{cases} \quad (3b)$$

There is no jump at $x = x_0$ and zero afterwards. To account for the possible electrostatic effects,[98] we additionally add the constant offset that we assume to be independent of concentration.

**Case II.** The coefficients $\beta < 0$, $\gamma > 0$ and $\delta = 0$ for the first order ferroelectrics of displacement type. For the case, the width of the PR loop is proportional to the spontaneous polarization $P_S = \sqrt{\frac{\sqrt{\beta^2 - 4\alpha\gamma} - \beta}{2\gamma}}$, the loop height is proportional to the coercive field $E_S = \frac{2}{5}(2\beta + \sqrt{9\beta^2 - 20\alpha\gamma})\left(\frac{2\alpha}{-3\beta - \sqrt{9\beta^2 - 20\alpha\gamma}}\right)^{3/2}$.[99] The PR loop area $S$, which has a quasi-rectangular shape in a quasi-static case, is roughly proportional to their product, and has sense only in the polar phase:

$$S = \begin{cases} S_0(2\beta + \sqrt{9\beta^2 - 20\alpha\gamma})\left(\frac{-2\alpha}{3\beta + \sqrt{9\beta^2 - 20\alpha\gamma}}\right)^{\frac{3}{2}} \sqrt{\frac{\sqrt{\beta^2 - 4\alpha\gamma} - \beta}{2\gamma}}, & T \leq T_C\left(1 - \frac{x}{x_{cr}}\right) + \frac{\beta^2}{4\alpha_T\gamma}, \\ 0, & T \geq T_C\left(1 - \frac{x}{x_{cr}}\right) + \frac{\beta^2}{4\alpha_T\gamma}, \end{cases}$$

(3b)

where $\alpha(T, x)$ is a linear function of x according to Eq.(2). Equation (3b) is too cumbersome for the machine learning. Allowing for $P_S = \sqrt{\frac{-\beta}{2\gamma}}$ at the transition temperature, $T = T_C\left(1 - \frac{x}{x_{cr}}\right) + \frac{\beta^2}{4\alpha_T\gamma}$, it can be simplified further as:

$$S(T, x) \approx \begin{cases} S_0\left(\left(1 - \frac{x}{x_{cr}}\right) - \frac{T}{T_C}\right)^{\frac{5}{4}}, & T \leq T_C\left(1 - \frac{x}{x_{cr}}\right) + \theta, \\ 0, & T \geq T_C\left(1 - \frac{x}{x_{cr}}\right) + \theta, \end{cases} \quad (4a)$$

Here $S_0$ and $x_{cr}$ are fitting parameters, $\theta = \frac{\beta^2}{4\alpha_T\gamma}$ and $T_C$ are tabulated. Note the expression (4a) is valid near the transition point only.

Introducing the concentration $x_0$ for which phase transition happens at room temperature, $T_R = T_C\left(1 - \frac{x}{x_0}\right)$, Eq.(4a) can be rewritten as:

$$S(T_R, x) = \begin{cases} S_R\left(1 - \frac{x}{x_0}\right)^{\frac{5}{4}} + S_J, & x \leq x_0, \\ 0, & x > x_0. \end{cases} \quad (4b)$$

There is a jump of the height $S_J$ at $x = x_0$ and zero afterwards. Similar to case I, we add a constant concentration-independent offset to account for the possible electrostatic effects

**Case III.** For the order-disorder type ferroelectrics we can put $\alpha < 0$, $\beta = \gamma = 0$ and $\delta > 0$. Important than within the model $\alpha$ is x-dependent but should be temperature independent.

Moreover, we can assume that the unknown parameter $P_0$ can be x-dependent. Eq.(1b) can be rewritten as

$$\alpha(x)P_3 + k_B T \, arctanh\left(\frac{P_3}{P_0}\right) - P_0 E_3 = 0. \qquad (4a)$$

Using a series expansion $arctanh(y) \approx y + \frac{1}{3}y^3 + \frac{1}{5}y^5$ in Eq.(4a) we can reduce it either to the Case I or to the Case II in dependence on the cutting term in the expansion. In the simplest case:

$$\alpha\left(\frac{P_3}{P_0}\right) + \left(\frac{k_B T}{P_0}\right)\left[\left(\frac{P_3}{P_0}\right) + \frac{1}{3}\left(\frac{P_3}{P_0}\right)^3\right] = E_3, \qquad (4b)$$

From Eq.(4b), the coercive field $E_c = \frac{2}{3\sqrt{3}}\sqrt{-\frac{3P_0}{k_B T}\left(\alpha + \frac{k_B T}{P_0}\right)^3}$, and the loop height is proportional to the spontaneous polarization $P_S = P_0\sqrt{-\frac{3P_0}{k_B T}\left(\alpha + \frac{k_B T}{P_0}\right)}$. The PR loop area $S$ is roughly proportional to their product, but has sense only when $\alpha + \frac{k_B T}{P_0} < 0$. So:

$$S(T, x) = \begin{cases} S_0 \frac{3P_0}{k_B T}\left(\alpha(x) + \frac{k_B T}{P_0}\right)^2, & \alpha(x) + \frac{k_B T}{P_0} < 0, \\ 0, & \alpha + \frac{k_B T}{P_0} \geq 0. \end{cases} \qquad (5a)$$

Here $S_0$, $P_0$ and $\alpha$ are fitting parameters, by the x-dependence can be assumed:

$$\alpha(x) = \alpha_0\left(1 - \frac{x}{x_{cr}}\right). \qquad (5b)$$

Note that this functional form is equivalent to case I, and hence these forms cannot be separated in this approximation.

**Table I.** Material parameters for bulk BFO at room temperature 293°K

| coefficient | BiFeO3 (collected and recalculated from Ref.[a]) |
|---|---|
| Symmetry | rhombohedral |
| $\varepsilon_b$ | 9 |
| $\alpha$ (×10$^7$C$^{-2}$·mJ) | −7.53, $T_C$ =1100 K, $\alpha_T$ = 8.37 10$^5$C$^{-2}$·mJ/K |
| $\beta$ (×10$^8$C$^{-4}$·m$^5$J) | $\beta_{11}$= 48, $\beta_{12}$= 8 |
| $\gamma$ (×10$^8$C$^{-6}$·m$^9$J) | *Poorly known, maybe zero* |

a) J. X. Zhang, Y. L. Li, Y. Wang, Z. K. Liu, L. Q. Chen, Y. H. Chu, F. Zavaliche, and R. Ramesh. J. Appl. Phys. 101, 114105 (2007).

### III. The MCMC diagnostics for hypoAL steps in Figure 7.

The effective number of samples (n_eff) and the Gelman-Rubin convergence criteria (r_hat) for the two hypoAL steps shown in Figure 7a and 7b are shown below. In each case the Bayesian inference was performed using NUTS algorithm with 5000 warmup steps (not to be confused with the warm-up phase in the hypoAL) and 5000 samples.

|  | Step 10 | |
|---|---|---|
|  | n_eff | r_hat |
| A | 52.86 | 1.02 |
| C_0 | 29.56 | 1.12 |
| C_1 | 23.90 | 1.01 |
| k_length[0] | 72.99 | 1.00 |
| k_scale | 74.53 | 1.05 |
| noise | 54.17 | 1.00 |
| x_c | 16.15 | 1.07 |

|  | Step 11 | |
|---|---|---|
|  | n_eff | r_hat |
| A | 68.99 | 1.00 |
| C_0 | 52.92 | 1.02 |
| C_1 | 26.52 | 1.01 |
| k_length[0] | 85.58 | 1.05 |
| k_scale | 85.73 | 1.02 |
| noise | 71.03 | 1.00 |
| x_c | 21.96 | 1.01 |

Similar to the example in Supplementary Note I, the problems with MCMC convergence may indicate that a model is only partially correct (for the completely wrong model, the GP kernel will take over resulting in a good convergence but a trivial solution).